\documentclass[12pt]{article}
  \usepackage{amsfonts}
  \usepackage{amsmath}
\usepackage{amssymb}
\usepackage{amscd}
\usepackage[dvips]{graphicx}
\begin{document}

~~
\bigskip
\bigskip
\begin{center}
{\Large {\bf{{{N-enlarged Galilei Hopf algebra and its twist deformations}}}}}
\end{center}
\bigskip
\bigskip
\bigskip
\begin{center}
{{\large ${\rm {Marcin\;Daszkiewicz}}$}}
\end{center}
\bigskip
\begin{center}
\bigskip

{ ${\rm{Institute\; of\; Theoretical\; Physics}}$}

{ ${\rm{ University\; of\; Wroclaw\; pl.\; Maxa\; Borna\; 9,\;
50-206\; Wroclaw,\; Poland}}$}

{ ${\rm{ e-mail:\; marcin@ift.uni.wroc.pl}}$}

\end{center}
\bigskip
\bigskip
\bigskip
\bigskip
\bigskip
\bigskip
\bigskip
\bigskip
\bigskip
\begin{abstract}
The N-enlarged Galilei Hopf algebra is constructed. Its twist deformations are considered and the corresponding twisted
space-times are derived.
\end{abstract}
\bigskip
\bigskip
\bigskip
\bigskip
\eject

\section{{{Introduction and preliminaries}}}

Presently, it is well known, that in accordance with the
Hopf-algebraic classification of all deformations of relativistic
and nonrelativistic symmetries, one can distinguish three
types of quantum spaces \cite{class1}, \cite{class2} (for details see also \cite{conservative}):\\
\\
{ \bf 1)} Canonical ($\theta^{\mu\nu}$-deformed) type of quantum space \cite{oeckl}-\cite{dasz1}
\begin{equation}
[\;\hat{ x}_{\mu},\hat{ x}_{\nu}\;] = i\theta_{\mu\nu}\;, \label{noncomm}
\end{equation}
\\
{ \bf 2)} Lie-algebraic modification of classical space-time \cite{dasz1}-\cite{lie1}
\begin{equation}
[\;\hat{ x}_{\mu},\hat{ x}_{\nu}\;] = i\theta_{\mu\nu}^{\rho}\hat{ x}_{\rho}\;,
\label{noncomm1}
\end{equation}
and\\
\\
{ \bf 3)} Quadratic deformation of Minkowski and Galilei  spaces \cite{dasz1}, \cite{lie1}-\cite{paolo}
\begin{equation}
[\;\hat{ x}_{\mu},\hat{ x}_{\nu}\;] = i\theta_{\mu\nu}^{\rho\tau}\hat{
x}_{\rho}\hat{ x}_{\tau}\;, \label{noncomm2}
\end{equation}
with coefficients $\theta_{\mu\nu}$, $\theta_{\mu\nu}^{\rho}$ and  $\theta_{\mu\nu}^{\rho\tau}$ being constants.\\
\\
Besides, it has been demonstrated in \cite{nh1}, \cite{nh2} and \cite{conservative} that in the case of
so-called standard, acceleration-enlarged and doubly enlarged Newton-Hooke Hopf algebras,
 the twist deformation
provides the new  space-time noncommutativity of the
form\footnote{$x_0 = ct$.},\footnote{ The discussed space-times have been  defined as the quantum
representation spaces, so-called Hopf modules (see \cite{oeckl}, \cite{chi},
\cite{bloch}, \cite{wess}), for quantum acceleration-enlarged
Newton-Hooke Hopf algebras.}
\begin{equation}
{ \bf 4)}\;\;\;\;\;\;\;\;\;[\;t,\hat{ x}_{i}\;] = 0\;\;\;,\;\;\; [\;\hat{ x}_{i},\hat{ x}_{j}\;] = 
if_{\pm}\left(\frac{t}{\tau}\right)\theta_{ij}(\hat{x})
\;, \label{nhspace}
\end{equation}
with time-dependent  functions
$$f_+\left(\frac{t}{\tau}\right) =
f\left(\sinh\left(\frac{t}{\tau}\right),\cosh\left(\frac{t}{\tau}\right)\right)\;\;\;,\;\;\;
f_-\left(\frac{t}{\tau}\right) =
f\left(\sin\left(\frac{t}{\tau}\right),\cos\left(\frac{t}{\tau}\right)\right)\;,$$
$\theta_{ij}(\hat{x}) \sim \theta_{ij} = {\rm const}$ or
$\theta_{ij}(\hat{x}) \sim \theta_{ij}^{k}\hat{x}_k$ and  $\tau$ denoting the time scale parameter. Moreover,
in $\tau$ approaching infinity limit there have been obtained the following  standard, acceleration-enlarged and doubly enlarged
 Galilei spaces \cite{nh1}, \cite{nh2}, \cite{conservative}
 \begin{equation}
{ \bf 5)}\;\;\;\;\;\;\;\;\;[\;\hat{ x}_{\mu},\hat{ x}_{\nu}\;] = i\alpha_{\mu\nu}^{\rho_1...\rho_n}\hat{
x}_{\rho_1}...\hat{ x}_{\rho_n}\;, \label{relation}
\end{equation}
 with $n = 0,1,\ldots,6$. It should be also noted that different relations  between all mentioned above quantum spaces-times ({\bf 1)}, { \bf 2)}, { \bf 3)}, { \bf 4)} and { \bf 5)}) have been summarized in paper \cite{conservative}.

In this article we construct the so-called N-enlarged Galilei Hopf algebra \;${\cal U}^{(N)}_0(\mathcal{G})$, which generates the following
transformations of classical nonrelativistic space
\begin{eqnarray}
&x_i& \longrightarrow \;\;\omega_{ij} x_j+ \sum_{n=0}^{N}a_{in}t^n\;,\label{trans1}\\
&t& \longrightarrow \;\;t+t_0\;, \label{trans2}
\end{eqnarray}
with $\omega_{ij}$ denoting the $O(d)$ rotations and with real parameters $a_{in}$ $(n=0,1,\ldots,N)$. Particularly, for $N=1,2,3$ the rules
(\ref{trans1}), (\ref{trans2}) correspond respectively to the standard, acceleration-enlarged \cite{luk1} and doubly enlarged \cite{luk2} Galilei Hopf symmetries. In the next step of our investigations we provide the (Abelian) twist deformations of \;${\cal U}^{(N)}_0(\mathcal{G})$ Hopf structure. Consequently,
we get the following two types of space-time noncommutativity
\begin{equation}
[\,t,\hat{x}_i\,]
=0\;\;\;,\;\;\;[\,\hat{x}_a,\hat{x}_b\,]
= i\alpha^{ij}t^{n+m}
 (\delta_{ai}\delta_{bj} -
\delta_{aj}\delta_{bi})\;,\label{type1}
\end{equation}
and
\begin{eqnarray}
[\,t,\hat{x}_i\,]
=0\;\;\;,\;\;\;
[\,\hat{x}_a,\hat{x}_b\,] =2i\alpha t^n
\left[\;\delta_{ia}(\hat{x}_k\delta_{bl} - \hat{x}_{l}\delta_{bk}) -
\delta_{ib}(\hat{x}_k\delta_{al} -
\hat{x}_{l}\delta_{ak})\;\right]\;,\label{type2}
\end{eqnarray}
which for $N=1,2,3$ reproduce the relation (\ref{relation}).

There are various motivations for present   studies. First of all, we construct explicitly the most general
Hopf algebra of nonrelativistic symmetries containing standard, acceleration-enlarged  and doubly enlarged  Galilei groups.
Secondly, we get the completely new quantum (twist-deformed) space-times associated with \;${\cal U}^{(N)}_{\alpha}(\mathcal{G})$ Hopf
structure. Such a result seems to be quite interesting due to the fact, that it extends in natural way the mentioned above classification
of the quantum spaces. Finally, it should be noted  that the obtained results permit to consider the classical as well as quantum particle models defined on new  noncommutative nonrelativistic space-times (\ref{type1}) and (\ref{type2}).

The paper is organized as follows. In first section we provide the N-enlarged Galilei Hopf algebra \;${\cal U}^{(N)}_0(\mathcal{G})$. The
second section is devoted to its twist deformations and to the derivation of corresponding quantum space-times. The comments on the N-enlarged Newton-Hooke
Hopf algebra and final remarks are presented in the last section.

\section{{{N-enlarged Galilei Hopf algebra \;${\cal U}^{(N)}_0(\mathcal{G})$}}}

In this section we construct the N-enlarged Galilei Hopf algebra which generates the space-time transformations (\ref{trans1}) and (\ref{trans2}).
First of all, it is easy to see that the corresponding generators are represented on the space of functions as follows
\begin{eqnarray}
M_{ij}&\rhd& f(t,\overline{x}) =i\left( x_{i }{\partial_j} -x_{j
}{\partial_i} \right) f(t,\overline{x})\;,\label{a1}\\
H&\rhd& f(t,\overline{x})=i{\partial_t}f(t,\overline{x})\;, \label{a2}\\
G_i^{(n)}&\rhd& f(t,\overline{x})=it^n
{\partial_i}f(t,\overline{x})\;\;\;;\;\;\;n=0,1,\ldots N\;,  \label{a3}
\end{eqnarray}
where $M_{ij}$, $H$, $G_i^{(0)} (=P_i)$ and $G_i^{(1)} (=K_i)$ can be identified with rotation, time translation, momentum and boost operators respectively. Next, by straightforward calculation one can find the following commutation relations
 \begin{eqnarray}
&&\left[\, M_{ij},M_{kl}\,\right] =i\left( \delta
_{il}\,M_{jk}-\delta _{jl}\,M_{ik}+\delta _{jk}M_{il}-\delta
_{ik}M_{jl}\right)\;\; \;, \;\;\; \left[\, H,M_{ij}\,\right] =0
 \;,  \label{q1} \\
&~~&  \cr &&\left[\, M_{ij},G_k^{(n)}\,\right] =i\left( \delta
_{jk}\,G_i^{(n)}-\delta _{ik}\,G_j^{(n)}\right)\;\; \;, \;\;\;\left[
\,G_i^{(n)},G_j^{(m)}\,\right] =0 \;,\label{q2}
\\
&~~&  \cr &&\left[ \,G_i^{(n)},H\,\right] =-inG_i^{(n-1)}\;,\label{q3}
\end{eqnarray}
which together with classical coproduct and antipode
\begin{eqnarray}
\Delta_0(a) = a\otimes 1 +1\otimes a\;\;\;,\;\;\;S_{0}(a) =-a\;,\label{cop}
\end{eqnarray}
define the N-enlarged Galilei Hopf algebra \;${\cal U}^{(N)}_0(\mathcal{G})$. It should be noted, that for $N=1,2,3$ we  get
the standard \;${\cal U}_0(\mathcal{G})$, acceleration-enlarged \;${\cal U}_0(\widehat{\mathcal{G}})$ and doubly enlarged  \;${\cal U}_0(\widehat{\widehat{\mathcal{G}}})$ Galilei Hopf structures proposed in \cite{luk1} and \cite{luk2} respectively.

\section{Twist deformations of N-enlarged Galilei Hopf algebra and the corresponding quantum space-times}

Let us now turn to the twist deformations of the Hopf structure provided in pervious section. First of all, in accordance with Drinfeld  twist procedure
\cite{twist1}-\cite{twist3}, the algebraic sector of twisted
N-enlarged Galilei Hopf algebra $\;{\cal U}^{(N)}_{\alpha}(\mathcal{G})$ remains
undeformed  (see (\ref{q1})-(\ref{q3})), while
the   coproducts and antipodes  transform as follows (see formula (\ref{cop}))
\begin{equation}
\Delta _{0}(a) \to \Delta _{\alpha }(a) = \mathcal{F}_{\alpha }\circ
\,\Delta _{0}(a)\,\circ \mathcal{F}_{\alpha }^{-1}\;\;\;,\;\;\;
S_{\alpha}(a) =u_{\alpha }\,S_{0}(a)\,u^{-1}_{\alpha }\;,\label{fs}
\end{equation}
with $u_{\alpha }=\sum f_{(1)}S_0(f_{(2)})$ (we use Sweedler's notation
$\mathcal{F}_{\alpha }=\sum f_{(1)}\otimes f_{(2)}$).
Besides, it should be noted, that the twist factor
$\mathcal{F}_{\alpha } \in {\cal U}^{(N)}_{\alpha}(\mathcal{G}) \otimes
{\cal U}^{(N)}_{\alpha}(\mathcal{G})$
satisfies  the classical cocycle condition
\begin{equation}
{\mathcal F}_{{\alpha }12} \cdot(\Delta_{0} \otimes 1) ~{\cal
F}_{\alpha } = {\mathcal F}_{{\alpha }23} \cdot(1\otimes \Delta_{0})
~{\mathcal F}_{{\alpha }}\;, \label{cocyclef}
\end{equation}
and the normalization condition
\begin{equation}
(\epsilon \otimes 1)~{\cal F}_{{\alpha }} = (1 \otimes
\epsilon)~{\cal F}_{{\alpha }} = 1\;, \label{normalizationhh}
\end{equation}
with ${\cal F}_{{\alpha }12} = {\cal F}_{{\alpha }}\otimes 1$ and
${\cal F}_{{\alpha }23} = 1 \otimes {\cal F}_{{\alpha }}$.

It is well known, that the twisted algebra $\;{\cal U}^{(N)}_{\alpha}(\mathcal{G})$ can be described in terms of
so-called classical $r$-matrix $r\in {\cal U}^{(N)}_{\alpha}(\mathcal{G}) \otimes {\cal U}^{(N)}_{\alpha}(\mathcal{G})$,
which satisfies the  classical Yang-Baxter equation (CYBE)
\begin{equation}
[[\;r_{\alpha},r_{\alpha}\;] ] = [\;r_{\alpha 12},r_{\alpha13} +
r_{\alpha 23}\;] + [\;r_{\alpha 13}, r_{\alpha 23}\;] = 0\;,
\label{cybe}
\end{equation}
where   symbol $[[\;\cdot,\cdot\;]]$ denotes the Schouten bracket
and for $r = \sum_{i}a_i\otimes b_i$
$$r_{ 12} = \sum_{i}a_i\otimes b_i\otimes 1\;\;,\;\;r_{ 13} = \sum_{i}a_i\otimes 1\otimes b_i\;\;,\;\;
r_{ 23} = \sum_{i}1\otimes a_i\otimes b_i\;.$$

In this article we consider two types of Abelian twist deformation of N-enlarged Galilei Hopf algebra, described by the following $r$-matrices\footnote{$a \wedge b = a \otimes b - b\otimes a.$}
\begin{eqnarray}
r^{(n,m)} &=&  \frac{1}{2}{\alpha^{ij}} G_i^{(n)} \wedge
G_i^{(m)}\;\;\;\;\;\;\;\, [\;\alpha^{ij} = -\alpha^{ji}\;]\;,
\label{rmacierze01}
\end{eqnarray}
and
\begin{eqnarray}
r^{(n)} &=&  \alpha G_i^{(n)}
\wedge M_{kl}\;\;\; [\;i,k,l - {\rm fixed},\;\;i \neq
k,l\;]\;,\label{rmacierzen02}
\end{eqnarray}
where $\alpha^{ij}$, $\alpha$ denote the deformation parameters.
Due to the Abelian character of the above carriers (all of them contain
 the mutually commuting elements of the algebra), the
corresponding twist factors can be obtained in a  standard way
\cite{twist1}-\cite{twist3}, i.e. they take the form
\begin{eqnarray}
{\cal F}^{{(n,m)}} = \exp
\left(ir^{(n,m)}\right)\;\;\;\;{\rm and}\;\;\;\;{\cal F}^{{(n)}} = \exp
\left(ir^{(n)}\right)\;.
\label{factors}
\end{eqnarray}
Of course, for $N=1,2,3$ we obtain the twist factors for \;${\cal U}_{\alpha}({{\mathcal{G}}})$, \;${\cal U}_{\alpha}({\widehat{\mathcal{G}}})$
and \;${\cal U}_{\alpha}(\widehat{\widehat{\mathcal{G}}})$ Hopf structures discussed in articles \cite{nh1}, \cite{nh2} and \cite{conservative}.

The corresponding quantum space-times are defined as the representation spaces (Hopf modules) for N-enlarged Galilei Hopf algebra
\;${\cal U}_{\alpha}^{(N)}({{\mathcal{G}}})$, with action of the generators $M_{ij}$, $H$ and $G_i^{(n)}$ given by (\ref{a1})-(\ref{a3})
(see \cite{oeckl}, \cite{chi}, \cite{bloch}, \cite{wess}). Besides, the $\star$-multiplication of arbitrary two functions covariant under $\;{\cal U}^{(N)}_{\alpha}(\mathcal{G})$ is defined as follows
\begin{equation}
f(t,\overline{x})\star^{(n,m)} g(t,\overline{x}):=
\omega\circ\left(
 (\mathcal{F}^{(n,m)})^{-1}\rhd  f(t,\overline{x})\otimes g(t,\overline{x})\right)
 \;,
\label{star1}
\end{equation}
\begin{equation}
f(t,\overline{x})\star^{(n)} g(t,\overline{x}):=
\omega\circ\left(
 (\mathcal{F}^{(n)})^{-1}\rhd  f(t,\overline{x})\otimes g(t,\overline{x})\right)
 \;,
\label{star2}
\end{equation}
where symbols  $\mathcal{F}^{(n,m)}$, $\mathcal{F}^{(n)}$ denote the  twist factors (see
(\ref{factors}))   and $\omega\circ\left( a\otimes b\right) =
a\cdot b$. Consequently, we get
\begin{equation}
[\,t,x_a\,]_{{\star}^{(n,m)}} =0\;\;\;,\;\;\;
[\,x_a,x_b\,]_{{\star}^{(n,m)}}
= i\alpha^{ij}t^{n+m}
 (\delta_{ai}\delta_{bj} -
\delta_{aj}\delta_{bi})\;,\label{rspacetime1}
\end{equation}
and
\begin{eqnarray}
[\,t,x_a\,]_{{\star}^{(n)}} =0\;\;\;,\;\;\;
[\,x_a,x_b\,]_{{\star}^{(n)}} =2i\alpha t^n
\left[\;\delta_{ia}(x_k\delta_{bl} - x_{l}\delta_{bk}) -
\delta_{ib}(x_k\delta_{al} -
x_{l}\delta_{ak})\;\right]\;,\label{rspacetime5}
\end{eqnarray}
respectively. Obviously, for $N=1,2,3$ we obtain the quantum spaces provided in \cite{nh1}, \cite{nh2} and \cite{conservative}. It should be also
 noted that for   deformation parameters $\alpha^{ij}$ and $\alpha$ approaching
zero, the above quantum  space-times  become classical.


\section{Comments on the N-enlarged Newton-Hooke Hopf algebra and final remarks}

In this short paper we provide the (new) N-enlarged Galilei Hopf structure $\;{\cal U}^{(N)}_{0}(\mathcal{G})$. Besides, we discuss its twist deformations and derive the corresponding quantum space-times. However, it should be noted that there still remain  the few open problems. Firstly, one can ask
about the relativistic counterpart of $\;{\cal U}^{(N)}_{0}(\mathcal{G})$ quantum group related by contraction scheme, which connects both relativistic and nonrelativistic  Hopf algebras. Moreover,  one should better understand the meaning of the parameters $a_{in}$ in the formula (\ref{trans1}). This aim can be achieved  by the analyzing  of
the classical and quantum dynamical models defined on the twisted spaces (\ref{type1}) and (\ref{type2}).
Finally, it seems to be quite interesting
to find the corresponding N-enlarged Newton-Hooke transformations, which in the limit of cosmological constant $(\tau)$ approaching infinity  reproduce
the rules (\ref{trans1}) and (\ref{trans2}). Unfortunately, the solution of such a problem seems to be calculationally   nontrivial  for arbitrary value of $N$. For example, if $N = 6$ one can derive\footnote{For $N=1,2,3$ the above transformations contain the symmetries proposed in \cite{bacry}, \cite{luk1}
and \cite{luk2} respectively.}
\begin{eqnarray}
&x_i& \longrightarrow \;\;\omega_{ij} x_j+ a_{i0} C_{\pm} \left(\frac{t}{\tau}\right) + a_{i1}  \tau S_{\pm} \left( \frac{t}{\tau}\right) \pm 2 a_{i2}  \tau^2
\left(C_{\pm} \left(\frac{t}{\tau}\right)  - 1\right) + \nonumber\\
&~~&~~~~~~\,\pm\;6a_{i3}\tau^3\left(S_{\pm} \left(\frac{t}{\tau}\right)  - \frac{t}{\tau}\right)
+
24a_{i4}\tau^4 \left(C_{\pm} \left(\frac{t}{\tau}\right)  \mp \frac{1}{2}\left(\frac{t}{\tau}\right)^2 - 1\right)
+\label{nhtrans1}\\
&~~&~~~~~~\,+120a_{i5}\tau^5
\left(S_{\pm} \left(\frac{t}{\tau}\right)  \mp \frac{1}{6}\left(\frac{t}{\tau}\right)^3 - \frac{t}{\tau}\right) + \nonumber\\
&~~&~~~~~~\,+720a_{i6}\tau^6\left(\pm C_{\pm} \left(\frac{t}{\tau}\right)  \mp \frac{1}{24}\left(\frac{t}{\tau}\right)^4 - \frac{1}{2}
\left(\frac{t}{\tau}\right)^2 \mp 1\right)\;, \nonumber\\
&t& \longrightarrow \;\;t+t_0\;, \label{nhtrans2}
\end{eqnarray}
with $C_{+} [\frac{t}{\tau}] = \cosh \left[\frac{t}{\tau}\right]$, $S_{+}
[\frac{t}{\tau}] = \sinh \left[\frac{t}{\tau}\right]$ in the case of "de-Sitter" as well as with
$C_{-} [\frac{t}{\tau}] = \cos \left[\frac{t}{\tau}\right]$, $S_{-}
[\frac{t}{\tau}] = \sin \left[\frac{t}{\tau}\right]$ in the case of "anti-de-Sitter" algebra. One can also check that in such a situation the corresponding
Hopf algebra is given by the commutation relations (\ref{q1})-(\ref{q3}) with additional relation $\left[\, H,G_i^{(0)}(P_i)\,\right] =\pm \frac{i}{\tau}G_i^{(1)}(K_i)$
and classical coproduct (\ref{cop}). The proper (N=)6-enlarged Newton-Hooke twisted spaces can be calculated as well, and, for example, one of  them takes the form
\begin{eqnarray}
[\,t,x_a\,]_{{\star}^{(6,6)}} &=& 0\;, \cr
[\,x_a,x_b\,]_{{\star}^{(6,6)}} &=& i\alpha^{ij}
518400\tau^{12}\left(\pm C_{\pm} \left(\frac{t}{\tau}\right)  \mp \frac{1}{24}\left(\frac{t}{\tau}\right)^4 - \frac{1}{2}
\left(\frac{t}{\tau}\right)^2 \mp 1\right)^2 \times \\ \nonumber
&\times& (\delta_{ai}\delta_{bj} -
\delta_{aj}\delta_{bi})\;.\label{nhhspacetime1}
\end{eqnarray}
The work in the mentioned above directions already started and are in progress.

\section*{Acknowledgments}
The author would like to thank J. Lukierski and A. Borowiec
for valuable discussions. This paper has been financially  supported  by Polish
NCN grant No 2011/01/B/ST2/03354.

\end{document}